# Excessive use, ill use and misuse of Bibliometrics

Rajeeva Laxman Karandikar
rkarandikar@gmail.com, rlk@cmi.ac.in

Professor Emeritus, Chennai Mathematical Institute
H1 Sipcot IT Park, Siruseri, Kelambakkam 603103
and
Chairperson, National Statistical Commission
Government of India.

Abstract

*Impact factor, H-index, citation index,* and such other indices have been playing an increasing role in scientific assessment of institutions, researchers, allocation of research funds,... across the globe. These indicies do not have any statistical basis but lots of decisions such as ranking of institutions, ranking of departments, assessment of faculty members for hiring and for promotions as well as selection for various awards are being made by using these indices. Several experts across disciplines have been writing that these indicies should have a marginal role, if any, and judgements should be based on critical assessment of the content by experts. But the dependence on these s steadily increasing. This article cites various such documents being published across the globe and across the disciplies and urging decision makers to ignore or at best give a minimal weight to such indices.

Over the last few decades, *Impact factor, H-index, citation index, ..* have been playing an increasing role in scientific assessment of institutions, researchers, allocation of research funds,... across the globe. Together, these *tools* are called *Bibliometrics*. Bibliometrics developed with a notable lack of explicit statistical foundation or serious discussion of its underlying principles. Given that the indices inherently rely on statistical concepts (e.g., counting citations, publications and combining these to rank individual researchers, institutions), numerous mathematicians and statisticians have critically examined the field over the last three decades, raising persistent questions about its validity and appropriateness as a measurement tool for academic purposes.

Science Citation Index (SCI) was introduced by Eugene Garfield in 1964, who founded *The Institute for Scientific Information (ISI)* as a private company and was distributed in printed form. In the 1970s, ISI introduced Social Science Citation Index (SSCI) and Arts & Humanities Citation Index (AHCI), Journal Citation Reports (JCR) (later renamed as Journal Impact Factor). ISI was acquired by *Thomson Scientific & Healthcare* later became Thomson Reuters.

Thomson Reuters made it available online in 1997. Later, it sold its Intellectual Property & Science business — which included the Institute for Scientific Information (ISI), Web of Science, and related bibliometrics products — for US $3.55 billion in cash in 2016.

It should be noted that while ISI has scholarly branding and historical prestige, it is not an independent academic or non-profit scientific body. Its products and rankings are part of a commercial analytics business.

Bibliometrics developed without any statistical foundation or serious discussion of its underlying principles. Given that it inherently relies on statistical concepts (e.g., counting citations, publications and combining these to rank individual researchers, institutions), numerous mathematicians and statisticians have critically examined the field over the last three decades, raising persistent questions about its validity and appropriateness as a measurement tool for academic purposes.

Doubts have been raised about the impact the excessive reliance on the bibliometrics are having by academicians is various disciplines. Here we will outline some significant writings in last 2 decades.

International Mathematical Union (IMU) in cooperation with the International Council of Industrial and Applied Mathematics (ICIAM) and the Institute of Mathematical Statistics (IMS) had constituted a committee of experts consisting of Robert Adler, John Ewing, and Peter Taylor, to examine the effects that quantitative assessment of research was having on their disciplines. The report of this committee [1] appeared in 2009 pointing out various pitfalls in relying on these indices in judging science. The report becomes even more relevant now as instead of curbing the widespread ill-use and misuse of citation data over the last 17 years, it has grown beyond what the authors of the report had imagined then, leading to what can be called *large scale gaming of quantitative metrics.*

The San Francisco Declaration on Research Assessment (DORA) [2] emerged during the annual meeting of the American Society for Cell Biology (ASCB) held in December 2012 in San Francisco. This came as a response to growing dissatisfaction with the excessive reliance on journal-based metrics — especially the Journal Impact Factor — in hiring, promotion, funding, and institutional evaluation. The declaration argued that the quality of individual research articles cannot be accurately judged by the citation performance or prestige of the journal in which they appear. DORA called upon universities, funding agencies, publishers, and researchers to adopt broader and more responsible methods of assessment that emphasize the intrinsic merit, rigor, openness, and societal contribution of scholarly work rather than narrow numerical indicators.

DORA rapidly became one of the most influential reform movements in global academia and research policy. Thousands of organizations (including Indian National Science Academy (INSA)) and tens of thousands of individual researchers across disciplines have endorsed it, including universities, scientific societies, and funding bodies. The declaration encouraged practices such as assessing research on its own content, recognizing diverse forms of scholarship including datasets and software, and reducing the misuse of bibliometric indicators in academic decision-making. Its influence has extended well beyond the natural sciences into the social sciences and humanities, where concerns about citation metrics and ranking systems are especially acute. DORA also helped catalyze wider international debates about research culture, responsible metrics, open science, and the commercialization of scholarly evaluation.

Bruce Alberts, [3] Editor-in-Chief of Science, writes in his editorial in Science about *The San Francisco Declaration on Research Assessment (DORA)* that the declaration states that *the impact factor must not be used as a surrogate measure of the quality of individual research articles, to assess an individual scientist's contributions, or in hiring, promotion, or funding decisions.*

Prof David Magnus, Professor of Medicine and Biomedical Ethics at Stanford Univer sity [4] writes an editorial in *The American Journal of Bioethics* titled *Overthrowing the Tyranny of the Journal Impact Factor* - the title itself indicates his view. He characterizes this over reliance as a "tyranny" because it distorts incentives across the research ecosystem. Scholars are pressured to target high-impact journals rather than pursue important or innovative work; journals may engage in practices that artificially boost citations; and institutions use JIF-driven metrics in hiring, promotion, and funding decisions. The result is a system that privileges prestige over substance, encourages strategic behavior ("gaming" the metric), and disadvantages fields or topics with naturally lower citation rates.

Various groups of academicians, cutting across subjects, have been writing for long about how these tools are having a negative impact on academics in general. More specifically, Prof Randy Wayne Schekman, University of California, Berkeley, who had been editor for various leading journals and shared the 2013 Nobel Prize for Physiology or Medicine, had written [5] a long article over a decade ago titled *How journals like Nature, Cell and Science are damaging science*, where among other things he says *A paper can become highly cited because it is good science – or because it is eye-catching, provocative or wrong. Luxury-journal editors know this, so they accept papers that will make waves because they explore sexy subjects.*

Yves Gingras, Canada Research Chair in the history and sociology of science at the University of Quebec in Montreal [6], has written a book *Bibliometrics and Research Evaluation: Uses and Abuses* in 2016. In this book, he offers a spirited argument against an unquestioning reliance on bibliometrics as an indicator of research quality. Gingras shows that bibliometric rankings have no real scientific validity, rarely measuring what they pretend to.

Stephen Curry, a professor of structural biology and assistant provost for equality, diversity and inclusion at Imperial College London was Chair of DORA steering group. In 2018, he wrote in Nature [7] that *while the San Francisco Declaration on Research Assessment (DORA) was a vital "rallying point" in 2012, it is no longer enough to sim ply sign a document. The "idealism" of the declaration must be translated into practical tools for fair evaluation. He notes that while momentum is building (particularly in the UK), the "familiar harness" of the Journal Impact Factor (JIF) remains difficult to root out.*
Committee on Publication Ethics (COPE), that has academicians from diverse back ground, has made similar observations n document "Citation Manipulation" (July 2019) [8] examines the growing problem of unethical citation practices driven by the increasing importance of bibliometric indicators such as citation counts and journal impact factors. COPE recommends that journals and publishers adopt clear policies regarding acceptable self-citation practices, educate editors and reviewers about ethical citation behaviour, and establish transparent procedures for addressing suspected manipulation. The document also notes that excessive dependence on citation-based metrics has created systemic incentives for unethical conduct, especially in highly competitive "publish or perish" academic environments.

*Gaming the Metrics: Misconduct and Manipulation in Academic Research* [9] is a major interdisciplinary critique of the modern "audit culture" in academia and the growing dependence on bibliometric indicators such as citation counts, impact factors, rankings, downloads, and altmetrics. Edited by Mario Biagioli and Alexandra Lippman, the volume argues that contemporary academia has shifted from "publish or perish" to "impact or perish," where scholarly success is increasingly measured through numerical indicators rather than substantive intellectual contribution. The contributors show that once metrics become targets for evaluation, they invite strategic manipulation — an illustration of Goodhart's Law: *when a measure becomes a target, it ceases to be a good measure.*

The book examines how metric-driven systems have transformed the ecology of aca demic research and publishing. Several chapters analyze practices such as citation cartels, coercive citation, peer-review manipulation, predatory publishing, fabricated editorial boards, and “salami slicing” of research into multiple minimal publications. Other essays explore how journal impact factors and university rankings reshape institutional behavior, encouraging universities, journals, publishers, and researchers to optimize for measurable visibility rather than genuine scholarly quality. The volume also studies the emergence of new forms of surveillance and forensic scrutiny through platforms like PubPeer and Retraction Watch, which attempt to detect manipulation and misconduct in real time.

A recurring theme in the book is that many problematic practices are not merely individual ethical failures but structural consequences of evaluation systems built around quantification. The contributors argue that excessive reliance on metrics creates incentives for gaming behaviour even among otherwise honest researchers. The book therefore situates academic misconduct within broader political and economic transformations in higher education, including managerialism, global competition, commercialization of publishing, and performance-based funding systems. Importantly, the volume also highlights global inequalities, showing how metric regimes can disadvantage scholars from the Global South while simultaneously creating markets for counterfeit journals, fake conferences, and predatory publication networks.

Rather than rejecting evaluation altogether, the book calls for a more reflective and context-sensitive approach to research assessment. Many contributors emphasize the need to combine quantitative indicators with qualitative peer judgment and disciplinary understanding. The volume has become an influential reference in debates surrounding responsible metrics, research integrity, publication ethics, and reforms associated with initiatives such as the (DORA). It is particularly important because it moves beyond narrow discussions of fraud and instead examines how entire systems of measurement reshape academic behaviour, institutional priorities, and even the meaning of scholarly excellence itself.

Clarivate, the company that is the prime mover of Bibliometrics (originally a part of Thomson Reuters, which sold it to private investors for a huge sum). In November 2023 Clarivate announced that it had excluded the entire field of mathematics from the most recent edition of its influential list of highly-cited authors. Explaining this decision, David Pendlebury, Head of Research Analysis at the Institute for Scientific Information wrote [10] *“We have excluded the Mathematics category from our analysis for 2023. The field of Mathematics differs from other categories in Essential Science Indicators (ESI). It is a highly fractionated research domain, with few individuals working on a number of specialty topics. The average rate of publication and citation in Mathematics is relatively low, so small increases in publication and citation tend to distort the representation and analysis of the overall field. Because of this, the field of Mathematics is more vulnerable to strategies to optimize status and rewards through publication and citation manipulation.*

*The responsible approach now is to afford this category additional analysis and judgement to*

*identify individuals with significant and broad influence in the field. We are consulting with leading bibliometricians and mathematicians to discuss our future ap proach to the analysis of this field."*

It should be noted that Clarivate is conceding that such things happen, and that the field of mathematics is more vulnerable for this. The company does not even say that this happens only in mathematics. It can be interpreted as Clarivate conceding that other sciences are also vulnerable. Perhaps it was clear to them that mathematicians can easily bring evidence of wrong doing (by users) and that would harm its business!

Most academicians may not have heard words - *Citation Cartels*, *Paper Mills*, *Citation broker*, *Hijacked journal*, *Predatory journal*, *Predatory conference*. Readers can do a google search with these terms and see the prevalence of these practices.

In spite of all these warnings by academic communities across the globe, the ill use and misuse of the indices has been growing at an alarming pace.

Recently, *the International Mathematical Union (IMU)* and *the International Council of Industrial and Applied Mathematics (ICIAM)* constituted a Working Group and it was given the task to explore the impact of bibliometrics on mathematics in current times. The report of the working group has just been published recently [12, 13]. The report is urging the mathematicians across the world to ignore (or at least give minimum weight) to any numbers, rankings etc of individual researchers, institutions produced via Bibliometrics when it comes to assessing researchers or institutions.

The extensive references given in [12] and the writings cited in this article have pointed out that various other subject experts are also worried about the same.

**Recommendation:**

**It is high time the decision makers, while ranking institutions and while choosing researchers for hiring, promotion or awards take note of this ill use and misuse of bibliometrics that is rampant and is growing. The least they can do is to significantly reduce the weight given to the "objective criterion" (using bibliometrics) and give significantly higher weight to "subjective criterion" (by choosing the experts suitably).**